\documentclass[pre]{revtex4}
\usepackage[cp1251]{inputenc}
\usepackage[english,russian]{babel}
\usepackage{graphicx}
\usepackage{dcolumn}
\usepackage{eucal}
\usepackage[dvips]{epsfig}
\usepackage{amssymb}
\usepackage{amsmath}

\begin{document}%\topmargin -0.5 in

%\linespread {1.6}

\newcommand{\bs}{\boldsymbol}
\newcommand{\mbb}{\mathbb}
\newcommand{\mcal}{\mathcal}
\newcommand{\mfr}{\mathfrak}
\newcommand{\mrm}{\mathrm}

\newcommand{\ovl}{\overline}
\newcommand{\p}{\partial}

\renewcommand{\d}{\mrm{d}}
\newcommand{\lap}{\triangle}

\newcommand{\lan}{\bigl\langle}
\newcommand{\ran}{\bigl\rangle}

\newcommand{\bse}{\begin{subequations}}
\newcommand{\ese}{\end{subequations}}

\newcommand{\be}{\begin{eqnarray}}
\newcommand{\ee}{\end{eqnarray}}

\newcommand{\ga}{\alpha}
\newcommand{\gb}{\beta}
\newcommand{\gc}{\gamma}
\newcommand{\gd}{\delta}
\newcommand{\gr}{\rho}
\newcommand{\eps}{\epsilon}
\newcommand{\veps}{\varepsilon}
\newcommand{\gs}{\sigma}
\newcommand{\gf}{\varphi}
\newcommand{\go}{\omega}
\newcommand{\gl}{\lambda}

\renewcommand{\l}{\left}
\renewcommand{\r}{\right}

%\draft
\title{\bf Coexistence of the "bogolons"\, and the one-particle spectrum of excitations with a gap in the degenerated Bose gas
}
\author{V.B. Bobrov, S.A. Trigger, I.M. Yurin}
\address{Joint\, Institute\, for\, High\, Temperatures, Russian\, Academy\,
of\, Sciences, 13/19, Izhorskaia Str., Moscow\, 125412, Russia;\\
email:\,satron@mail.ru}

\begin{abstract}
Properties of the weakly non-ideal Bose gas are considered without
suggestion on C-number representation of the creation and
annihilation operators with zero momentum. The "density-density"\,
correlation function and the one-particle Green function of the
degenerated Bose gas are calculated on the basis of the
self-consistent Hartree-Fock approximation. It is shown that the
spectrum of the one-particle excitations possesses a gap whose
value is connected with the density of particles in the
"condensate". At the same time, the pole in the
"density-density"\, Green function determines the phonon-roton
spectrum of excitations which exactly coincides with one
discovered by Bogolyubov for the collective excitations (the "bogolons").\\

PACS number(s): 05.30.Jp, 03.75.Kk, 03.75.Nt, 05.70.Fh\\

\end{abstract}

\maketitle

\section{Introduction}

It is well known that theoretical description of the superfluidity
phenomenon must take into account  the interaction between helium
atoms. Also, the closeness of the transition temperature in
superfluid state $T_\lambda$ to the temperature $T_0$ of Bose
condensation for the ideal Bose gas and accumulation of the
macroscopic quantity of particles in the state with the momentum
zero ("condensate") [1] are the most important features which
permit us to apply the models of weakly non-ideal Bose gas to the
liquid HeII. Therefore, it seems possible that the phenomenon of
superfluidity can be described within the framework of the weakly
non-ideal Bose gas model.

It is necessary to stress here that the model of the ideal Bose
gas does not satisfy the Landau criterion of superfluidity [2,3]
and cannot explain the specific behavior of the thermodynamic
properties of the superfluid helium [4]. However, the application
of the standard perturbation theory expansion on the
inter-particle interaction to the Bose gas at $T<T_0$ at once
faces the problem of the appropriate description of the condensate
(see, e.g., [5]).

 Therefore, the development of the microscopic theory of the
 degenerated Bose system can be conducted in two ways:

 A) The formulation  of some special suggestions (conditions) for
 some functions (operators) for the degenerated Bose gas,

 B) Using another initial model instead of the ideal
 Bose gas for developing the perturbation theory.

Starting with the classical Bogolyubov's papers [6,7], the
microscopic theory of the degenerated Bose gas has been based on
the special suggestion on the C-number representation of the
creation $a_0^{+}$ and annihilation $a_0$ operators of the
particles with the momentum ${\bf p}=0$.

The results obtained by Bogolyubov, including the spectrum of the
collective excitations (that further be called "bogolons"\, after
the name of the author), permit us to give the qualitative
explanation of the experimental data in superfluid helium and to
satisfy the Landau condition of superfluidity. It should be
mentioned here that the mathematical methods of the quantum field
theory were first applied to the study of the non-ideal
degenerated Bose gas by Belyaev [8], without special suggestion on
the C-number representation of the operators $a_0^{+}$ and $a_0$.
In [8], the special diagram technique for the perturbation row at
zero temperature, was developed and then generalized for the one-
particle Green functions. In particular, it was suggested to
consider (apart from the usual Green function with one incoming
and, respectively, one outgoing external lines), the additional
Green functions with two incoming and two outgoing external lines.
However, due to the complexity of its mathematical approach, the
method suggested in [8] is not widely used.

Hugenholtz and Pines in [9] reformulated the problem at the
beginning, by changing, according to Bogolyubov, the operators
$a_0^{+}$ and $a_0$ by C-numbers, and they could use almost
automatically the quantum field theory methods for investigating
the Bose gas with the "condensate". In particular, in [9], it was
shown that for the C-number representation of the operators
$a_0^{+}$ and $a_0$ in the excitation spectrum connected with
one-particle Green function, the gap cannot exist. Their approach
is used presently in the theory of the degenerated Bose gas (see,
e.g., [5]).

However, rigorous proof of the correctness of the C-number
representation of the operators $a_0^{+}$ and $a_0$ is not
possible and, due to this circumstance, the correspondence of the
initial Hamiltonian to the Hamiltonian which arises after changing
the operators $a_0^{+}$ and $a_0$ on C-numbers, remains undecided
[10,11]. Moreover, in [12-14], the attempts were made to suggest
some canonical transformation of the field operators, which would
not be connected with the C- number representation of the
operators $a_0^{+}$ and $a_0$ (and therefore alternative to the
Bogolyubov's one).

Importantly, in [13-16], the possibility of the existence of two
different spectra simultaneously - the one-particle one with a gap
and the collective one - corresponding to the Bogolyubov's branch
of excitations, has been suggested and discussed.

In addition, in [14,18], most of the known results for the
thermodynamic properties of the degenerated weakly non-ideal Bose
gas have been reproduced by using the "dielectric formalism"\,
without C-number representation of the operators  $a_0^{+}$ and
$a_0$. Furthermore, in [19], it was argued that the formal
restrictions on the use of the standard temperature diagram
technique for the temperatures $T<T_0$ are absent.

Another essential circumstance, connected with the C-number
representation of the operators  $a_0^{+}$ and $a_0$, has to be
mentioned. The C-number representation of these operators is based
on the commutation relations and the smallness of the parameter
$1/N_0$, where $N_0$ is the operator of the particle number in the
state ${\bf p}=0$. Therefore, this parameter makes sense only for
the system with fixed number of particles, i.e. in the framework
of the canonical ensemble where $N_0$ is equal to the average
$<N_0>$. Although the Bogolyubov's canonical transformation of the
initial Hamiltonian is realized in the canonical ensemble, the
calculations of the averages, usually are executed in the grand
canonical ensemble.

In the present paper  we are exploring the second way, B, for
developing the microscopic theory of the degenerated Bose system.
We are take into account not only the provision of the Landau
superfluidity condition, but also the consecutive description of
the weakly non-ideal Bose gas for the transition from temperatures
$T\geq T_0$ to the case of $T<T_0$ including the case of a strong
degeneration. As an initial approach, we consider the
self-consistent Hartree-Fock approximation, which is the best
one-particle approximation for the normal systems (see, e.g.,
[20]).

\section{Hartree-Fock approximation for $ {\bf \delta(r)}$ - potential
and the gap in the one-particle spectrum}

Let us consider the weakly non-ideal Bose gas which consists of
the particles with the mass $m$ and zero spin. The interaction
between the particles is described by the potential $U(r)$ with
the Fourier-component $u({\bf q})$
\begin{eqnarray}
\lim_{q\rightarrow 0} u(q)=u(0)\equiv u({\bf q}=0)>0 \label{A1}
\end{eqnarray}
Expression for the one-particle distribution function $f({\bf p})$
reads
\begin{eqnarray}
f({\bf p})=<a_{\bf p}^{+}a_{\bf p}>,\;\; n=\frac{1}{V}\sum_p
f({\bf p}), \label{A2}
\end{eqnarray}
where $a_{\bf p}^{+}$ and $a_{\bf p}$ are the creation and
annihilation operators for the particles with the momentum $\hbar
{\bf p}$, $n=<N>/V$ is the average density of the particles in the
volume $V$ at temperature $T$, $N=\sum_p a_{\bf p}^{+} a_{\bf p}$,
and the angle brackets <...> denote the averaging in the grand
canonical ensemble with the chemical potential $\mu$.

Now let us pay attention to the spectral representation which
shows that the function $f ({\bf p})$ is not negative for the
arbitrary wave vectors ${\bf p}$
\begin{eqnarray}
f({\bf p})\geq 0 \label{A3}
\end{eqnarray}
In the framework of the self-consistent Hartree-Fock approximation
for $T>T_0$, the distribution function $f({\bf p})$ satisfies the
relation (see,e.g., [21])
\begin{eqnarray}
f({\bf p})=\left\{exp\left(\frac{E({\bf
p})-\mu}{T}\right)-1\right\}^{-1} \label{A4}
\end{eqnarray}
where $E({\bf p})$ is the energy of the one-particle excitations
\begin{eqnarray}
E({\bf p})=\varepsilon(p)+nu(0)+\frac{1}{V}\sum_{{\bf q\neq
0}}u(q)f({\bf p+q})\label{A5}
\end{eqnarray}
$\varepsilon(p)=\hbar^2p^2/2m$ is the energy spectrum of a free
particle, the second and the third terms on the right side of
(\ref{A5}) correspond to the inter-particle interaction in the
Hartree and in the Fock (the so-called exchange interaction, which
is conditioned by the identity of the particles) approximations,
respectively.

As it will be clear from further consideration, the important
point is connected with the presence in (\ref{A5}) of the
distribution function $f({\bf p})$ determined by (\ref{A4}), but
not the distribution function $f^{id}({\bf p})$ of the ideal gas
with spectrum of the free particles, as it is usual in the
perturbational Hartree-Fock approximation. Therefore, equations
(\ref{A4}),(\ref{A5}) form the closed system of equations for
determination of the distribution function and the one-particle
excitation spectrum for the fixed thermodynamic parameters. In the
language of the temperature diagram technique [5,21], it means the
exact summation of some class of the diagrams in the equation for
the one-particle Green function.

For the particular case when the interaction potential has the
form of the $\delta$-potential
\begin{eqnarray}
U({\bf r})=u(0)\delta({\bf r}),\;\; u(q)=u(0)\label{A6}
\end{eqnarray}
from (\ref{A4}) and (\ref{A5}) we obtain
\begin{eqnarray}
f({\bf p})=\left\{exp\left(\frac{\varepsilon({\bf p})-\gamma({\bf
p})-\mu^{\ast}}{T}\right)-1\right\}^{-1} \label{A7}
\end{eqnarray}
\begin{eqnarray}
\mu^{\ast}=\mu-2nu(0), \,\,\, \gamma({\bf p})=\frac{u(0)}{V}f({\bf
p})\label{A8}
\end{eqnarray}
As is follows from Eqs.~(\ref{A7}),(\ref{A8}) before transition to
the thermodynamic limit, it is necessary to take into account that
the system is located in the finite volume $V$. The importance of
this point has been mentioned in [9] in the case of the ideal Bose
gas condensation.

Taking into account (\ref{A8}), one can see that
Eqs.~(\ref{A7}),(\ref{A8}) make sense only for $\mu^{\ast}<0$. As
this takes place, the contribution of the function $\gamma({\bf
p})$ in (\ref{A7}),(\ref{A8}) is negligible for the arbitrary
values of the vector ${\bf p}$. In this case, the distribution
function $f({\bf p})$ is equivalent to the one for the ideal Bose
gas $f^{id}({\bf p})$ with the chemical potential $\mu$ changed to
$\mu^{\ast}$. Correspondingly, in the transition from $\mu^{\ast}$
to the chemical potential $\mu^{id}$, all known results for the
ideal Bose gas are reproduced up to the temperature $T_0$ of Bose
condensation, including the transition temperature itself.

However, the situation changes dramatically for the temperature of
Bose condensation $T_0$, when the accumulation of the particles in
the state with the momentum ${\bf p}=0$ starts, and also below
this temperature.

By analogy with the case of the ideal Bose gas (see, e.g., [1]) it
would seem that one can suppose that
\begin{eqnarray}
\mu^{\ast}=0, \,\mbox{or}\,\mu=2nu(0),\label{A9}
\end{eqnarray}
and that the function $f({\bf p})$ can be represented in the form
\begin{eqnarray}
f({\bf p})=<N_0>\delta_{{\bf p},0}+f^T({\bf p})(1-\delta_{{\bf
p},0}), \label{A10}
\end{eqnarray}
where $N_0=a_0^{+}a_0$ is the operator of the quantity of
particles with the momentum equal zero ("condensate"), $f^T({\bf
p})$ is the one-particle distribution function with non-zero
momenta (the "overcondensate"\, states) and
\begin{eqnarray}
<N_0>=<N>\left\{1-\left(\frac{T}{T_0}\right)^{3/2}\right\}, \;\;
f^T({\bf p})=f_{id}^T ({\bf p})=
\left\{exp\left(\frac{\varepsilon({\bf
p})}{T}\right)-1\right\}^{-1}  \label{A11}
\end{eqnarray}
It is necessary to mention that the relation $f({\bf
p})=<N_0>\delta_{{\bf p},0}$ (see Eq.~(\ref{A10}) for $f^T({\bf
p})=0$) was first suggested in [22].

However, the representation (\ref{A9}) for the chemical potential
$\mu$ is twice as big as the chemical potential of the ideal Bose
gas in the case of strong degeneration. Besides, the function
$f_{id}^T ({\bf p})$ possesses the singularity at small wave
vectors ${\bf p}$.

Let us mention now that neglecting the value $\gamma({\bf p})$ at
${\bf p}=0$ for the temperatures $T<T_0$, is not correct for
calculating the distribution function $f({\bf p})$ in (\ref{A7})
\begin{eqnarray}
\gamma({\bf p})= n_0 u(0),\;\; \; n_0=\frac{<N_0>}{V}  \label{A12}
\end{eqnarray}
Therefore, if representation (\ref{A10}) is true, the
Eqs.~(\ref{A9}),(\ref{A11}) are wrong. Taking into account
(\ref{A10}),(\ref{A12}) from Eqs.~(\ref{A7}),(\ref{A8}) for
$T<T_0$ one directly finds
\begin{eqnarray}
\mu=(2n-n_0)u(0),\label{A13}
\end{eqnarray}
\begin{eqnarray}
f^T(p)=\left\{exp\left(\frac{E^{\ast}(
p)}{T}\right)-1\right\}^{-1}(1-\delta_{{\bf p},0}),\;\;\;E^{\ast}(
p)=\varepsilon(p)+n_0 u(0), \label{A14}
\end{eqnarray}
\begin{eqnarray}
n_0=n-\int \frac{d^3 p}{(2\pi)^3}f^T(p). \label{A15}
\end{eqnarray}
On the basis of (\ref{A13})-(\ref{A15}) it is easy to establish
the correctness of the following relations
\begin{eqnarray}
\lim_{T \rightarrow 0} n_0=n, \;\;\; \lim_{T \rightarrow
0}\mu=nu(0), \label{A16}
\end{eqnarray}
\begin{eqnarray}
\Delta=\lim_{p\rightarrow 0} E^{\ast}(p)=n_0 u(0), \label{A17}
\end{eqnarray}
\begin{eqnarray}
\lim_{p \rightarrow 0} f^T(p)= \left\{\exp
\left(\frac{\Delta}{T}\right)-1 \right\}^{-1}<\infty, \;\;\;
\lim_{T \rightarrow 0}f^T(p)=0. \label{A18}
\end{eqnarray}
Therefore, in the framework of the self-consistent Hartree-Fock
approximation (\ref{A4}),(\ref{A5}) for the one-particle
distribution function $f^T(p)$, for the temperatures $T<T_0$ we
can establish that:

(A) In the case of strong degeneration, the obtained results
(\ref{A16}) coincide with the known relations for the weakly
non-ideal Bose gas [5];

(B) In the one-particle excitation spectrum, the gap (\ref{A17})
arises between the "condensate"\, and the "overcondensate"\,
states. The appearance of the gap (\ref{A17})  is conditioned by
the existence of the "condensate". The spectrum of the
one-particle excitations, taking into account the "condensate"\,
with the zero energy and the spectrum for the "overcondensate"\,
states, satisfy the Landau condition of superfluidity;

(С) The distribution function for the "overcondensate"\, states is
finite at small values of the wave vectors distinct from the
distribution function (\ref{A11}) for the ideal Bose gas. On the
basis of the above, it is possible to assert that the
self-consistent Hartree-Fock approximation for the one-particle
distribution function (\ref{A4}),(\ref{A5}) is suitable as an
initial approximation for constructing the theory of Bose gas that
takes into account the interaction at arbitrary thermodynamic
parameters. It is necessary to mention further that according to
(\ref{A14}) in many applications, the condition $T\rightarrow 0$
is equivalent to the condition
\begin{eqnarray}
T\ll \Delta. \label{A19}
\end{eqnarray}

\section{Collective excitations and the dielectric formalism}

Having the above results, let us consider the problem of the
collective excitations in the weakly non-ideal Bose gas on the
basis of the "dielectric formalism" [17,18]. The experimental
determination of the collective excitations spectrum is
interpreted on the basis of the existing data on the well observed
maximums [23] in the dynamical structure factor $S(q,\omega)$ for
$q \neq 0$,
\begin{eqnarray}
S(q,\omega)=\frac{1}{V}\int^\infty_{-\infty}dt \,exp(i\omega
t)<\rho_q (t)\rho_ {-q} (0)>, \label{A20}
\end{eqnarray}
\begin{eqnarray}
\rho_q (t)= \sum_p \, a^{+}_{{\bf p-q}/2}(t)a_{{\bf p+q}/2}(t),
\label{A21}
\end{eqnarray}
where $\rho_q (t)$ is the Fourier-component of the operator of
particle density in the Heisenberg representation. The dynamical
structure factor (\ref{A20}) is directly connected [24] with the
retarded density-density Green function $\chi^R(q,z)$ which is
analytical in the upper semi-plane of the complex $z$ ($Im z>0$),
\begin{eqnarray}
S(q,\omega)=-\frac{2\hbar}{1- exp(-\hbar \omega/T)}\,Im
\chi^R(q,\omega+i0), \label{A22}
\end{eqnarray}
\begin{eqnarray}
\chi^R(q,z)=-\frac{i}{\hbar V}\int^\infty_0 dt \,exp(iz t)<[
\rho_q (t)\rho_{-q} (0)]>>=\frac{1}{V}<<\rho_q \mid
\rho_{-q}>>_{z}, \label{A23}
\end{eqnarray}
The equalities (\ref{A20}),(\ref{A22}) have to be taken in the
thermodynamic limit: $V \rightarrow\infty$, $<N>\rightarrow\infty$
and $<N>/V \rightarrow const$. Eq.~(\ref{A22}) serves as the basis
for calculation of the function $S(q,\omega)$ for quantum systems
by the perturbation methods of the diagram technique [5,21,24].
The retarded Green function $\chi^R(q,z)$ (\ref{A23}) is the
analytical continuation of the temperature Green function
$\chi^T(q,i\Omega_n)$
\begin{eqnarray}
\chi^T(q,i\Omega_n)=\frac{1}{V}<< \rho_q \mid
\rho_{-q}>>_{i\Omega_n}, \label{A24}
\end{eqnarray}
from the discrete multitude of the points on the imaginary axis
$i\Omega_n=i 2\pi n T$ to the upper semi-plane of the complex $z$
[5,21,24]. For the function $\chi^T(q,i\Omega_n)$, there exists
the diagram representation which is connected with the extraction
of the irreducible (in $q$-channel on one line of interaction
$u(q)$) part of $\chi^T(q,i\Omega_n)$ - the so-called the
polarization operator $\Pi(q,i\Omega_n)$ [21]. After analytical
continuation of the function $\chi^T(q,i\Omega_n)$, we arrive at
the expression
\begin{eqnarray}
\chi^R(q,z)=\frac{\Pi(q,z)}{\varepsilon(q,z)}. \label{A25}
\end{eqnarray}
Here, the function $\varepsilon(q,z)$, by analogy with the
terminology accepted in the theory of Coulomb systems [24], is
called dielectric permittivity
\begin{eqnarray}
\varepsilon(q,z)=1-u(q)\Pi(q,z). \label{A26}
\end{eqnarray}
It should be noted that all relations mentioned above in this
section are valid for arbitrary interaction $u({\bf q})$.

Determination of the appropriate approximation for the
polarization operator permits us to find the poles of the Green
function $\chi^R(q,z)$. These poles describe the collective
excitations in the system, which are the solutions of the equation
\begin{eqnarray}
\varepsilon(q,z)=0. \label{A27}
\end{eqnarray}
The above equation, meanwhile, is well known from the theory of
the Coulomb systems [25].

In considering the case of the weakly non-ideal Bose gas for
calculation of the polarization operator $\Pi(q,z)$, we restrict
ourselves to the simplest "one-loop"\, approximation, which in the
theory of the Coulomb systems [25] is called "random phase
approximation"\, (RPA). Then, taking into account Eq.~(\ref{A10})
one obtains [17,18]
\begin{eqnarray}
\Pi^{RPA}(q,z)=\Pi^{(0)}(q,z)+\Pi ^T(q,z). \label{A28}
\end{eqnarray}
\begin{eqnarray}
\Pi^{(0)}(q,z)=\frac{2n_0\varepsilon(q)}{\hbar^2z^2-\varepsilon^2(q)}
. \label{A29}
\end{eqnarray}
and
\begin{eqnarray}
\Pi^{T}(q,z)=\int\frac{d^3p}{(2\pi)^3}\, \frac{f_{id}^T({\bf
p-q}/2)-f_{id}^T({\bf p+q}/2)}{\hbar z +\varepsilon ({\bf
p-k}/2)-\varepsilon ({\bf p+k}/2)} \label{A30}
\end{eqnarray}

Taking into account the above consideration, we now modify the RPA
approximation on the generalized (MRPA) approximation, by
introducing the change in $\Pi^{T}(q,z)$ (\ref{A30}) operator. The
function $f_{id}^T$ (\ref{A11}) and the spectrum of the
one-particle excitations $\varepsilon (p)$ for the ideal Bose gas
are changing for the function $f^T(p)$ and the energy $E^\ast(p)$
(\ref{A14}) in the self-consistent Hartree-Fock approximation. As
this takes place, we observed that due to the ruptured character
of the one-particle spectrum at the point ${\bf p}=0$, the
function $\Pi^{(0)}(q,z)$ preserves its earlier form (\ref{A29}).
Then
\begin{eqnarray}
\Pi^{MRPA}(q,z)=\Pi^{(0)}(q,z)+\Pi_{MRPA}^T(q,z), \label{A31}
\end{eqnarray}
\begin{eqnarray}
\Pi_{MRPA}^{T}(q,z)=\int\frac{d^3p}{(2\pi)^3}\, \frac{f^T({\bf
p-q}/2)-f^T({\bf p+q}/2)}{\hbar z +E^\ast({\bf p-k}/2)-E^\ast({\bf
p+k}/2)}. \label{A32}
\end{eqnarray}

Considering further the case of a low temperature in (\ref{A19}),
one can omit the part $\Pi_{MRPA}^{T}(q,z)$ (\ref{A32}) in
$\Pi^{MRPA}(q,z)$. In this case, from (\ref{A25})-(\ref{A27}),
(\ref{A29}) it directly follows  that for the temperatures
$T\ll\Delta$
\begin{eqnarray}
\chi^R(q,z)=\frac{2n_0\varepsilon(q)}{\hbar^2z^2-(\hbar
\omega(q))^2}, \label{A33}
\end{eqnarray}
where the spectrum of the collective excitations is determined by
the equality
\begin{eqnarray}
\hbar \omega(q)=\left\{(\varepsilon(q))^2+2\varepsilon(q)n
u(q)\right\}^{1/2}. \label{A34}
\end{eqnarray}
The relation (\ref{A34}) completely coincides with the spectrum of
the "bogolons"\, for the interaction potential, where its
dependence on the wave vector [26,27] provides the fulfilment of
the Landau superfluidity condition. Since, in the temperature
interval under consideration, the density of the particles in the
"codensate"\, $n_0$ is close to the total density $n$, we can
change $n_0$ to $n$ in (\ref{A34}). Then, inserting (\ref{A33}) in
(\ref{A22}) and taking into account the determination of the
structure factor [1,23], we find [17,18]
\begin{eqnarray}
n S(q)=\frac{\varepsilon(q)}{\hbar\omega(q)}\,
cth\left\{\frac{\hbar\omega(q)}{2T}\right\} \label{A35}
\end{eqnarray}

The above equation is the generalization of the Feinman formula
[28] for the connection between the static structure factor and
the spectrum
\begin{eqnarray}
\hbar\omega(q)=\varepsilon(q)/S(q),\label{A36}
\end{eqnarray}
which is valid for the case $\hbar\omega(q)\gg T$.

In the opposite case of $\hbar\omega(q)\ll T$ (and $n_0\simeq n$,
e.g., $T \ll T_0$), we have
\begin{eqnarray}
S(q)=\frac{2\varepsilon(q)T}{\hbar^2\omega^2(q)},\;\; \lim_{q
\rightarrow 0} S(q)=\frac{T}{n u(0)} \label{A37}
\end{eqnarray}

Relation (\ref{A37}) corresponds to the general result for the
systems with the short-range interaction potential (1) [1]

\begin{eqnarray}
\lim_{q\rightarrow 0}S(q)=nT K_T,\;\;\;
K_T=-\frac{1}{V}\left(\frac{\partial V}{\partial P}\right)_T.
\label{A38}
\end{eqnarray}
Here $K_T$ is the isothermal compressibility of the system.
Comparing (\ref{A37}) and (\ref{A38}) and taking into account that
the spectrum of the "bogolons"\,  in the region of the small wave
vectors has the form

\begin{eqnarray}
\hbar\omega (q)=S_{T}\, q, \;\; S_T=\left(\frac{n
u(0)}{m}\right)^{1/2} \label{A39}
\end{eqnarray}
we conclude that the value $S_T$ characterizes the isothermal
sound velocity. In addition, on the basis of Eq.~(\ref{A35}) for
the static structure factor we can determine the free energy $F$
of weakly non-ideal Bose gas for the temperature $T\ll \Delta$ by
using the general relation [1]
\begin{eqnarray}
F=F^{id}+ \frac{1}{2}u(0) n N- \frac{1}{2} n \sum_q u(q) +
\frac{1}{2} n \sum_q \int_0^1 u(q) S_{\lambda}(q) d\lambda,
\label{A40}
\end{eqnarray}
\begin{eqnarray}
F^{id}=T \sum_q ln \left[1-\exp(-\varepsilon(q)/T)\right].
\label{A41}
\end{eqnarray}
Here $F^{id}$ is the free energy of the ideal Bose gas for $T<T_0$
[26] and $S_{\lambda}(q)$ is the static structure factor for the
system with the interaction potential $\lambda u(q)$. Inserting
(\ref{A35}) in (\ref{A40}) we find [17,18]
\begin{eqnarray}
F=\frac{1}{2}u(0) n N- \frac{1}{2} \sum_q \{\varepsilon(q)+n
u(q)-\left(\varepsilon(q)^2+2\varepsilon(q) n u(q)\right)^{1/2}\}
+ T \sum_q ln \left\{1-\exp[-\hbar\omega(q)/T] \right\}.
\label{A42}
\end{eqnarray}
Relation (\ref{A42}) completely corresponds to the results
obtained in [5,24,27].

\section{Conclusions}

In the present work we developed the theory of weakly non-ideal
Bose gas, below the condensation temperature, on the basis of the
self-consistent Hartree-Fock approximation. It was found that the
spectra of the collective excitations and the one-particle
excitations are distinct, which is contrary to the theory based on
the C-number representation for the operators $a_0^{+}$ and $a_0$.
It is shown that one-particle branch of excitations has a gap.
Both spectra satisfy the Landau criterion of superfluidity. It
must be noted that in the early works of Landau, and in the works
of Bogolyubov, they considered the possibility of the existence of
the gap in the spectrum of Bose system for $T<T_0$. However, they
omitted this suggestion because the phonon branch was absent.

It follows now from the present paper, that the collective
phonon-roton spectrum and the one-particle spectrum with a gap
actually coexist. On the basis of  the calculation of the
one-particle distribution function and "density-density"\, Green
function, in the framework of the self-consistent Hartree-Fock
approximation for the weakly non-ideal Bose gas, we can establish
the following:

1) This system possesses two branches of excitations - the
one-particle branch and the collective branch, each of them
satisfying the Landau condition of superfluidity.

2) In the region of small wave numbers, there is the gap in the
spectrum of the one-particle excitations which is conditioned by
the presence of the "condensate".

3) The spectrum of the "bogolons"\, corresponds to the
phonon-roton excitations that are observed in the experiments on
the neutron nonelastic collisions [29,30].

Therefore, even a weak inter-particle interaction leads to the
drastic differences in the description of the one-particle
distribution function and of the excitations of the
"overcondensed"\, particles. We have emphasizes that (as in the
case of C-number representation of the operators $a_0^{+}$ and
$a_0$) the application of the Hartree-Fock approximation to the
calculation of the Green functions for Bose gas in the temperature
region $T<T_0$ cannot be absolutely rigorously theoretically
justified and there is a need here for further experimental
examination. The principal difference between the results of this
paper and the traditional approach based on the C-number
representation for $a_0^{+}$ and $a_0$, is the appearance of the
gap in the spectrum of the one-particle excitations. It follows
from the results of this work that the gap cannot manifest itself
in the "density-density"\, Green function (at least in the
Hartree-Fock approximation) and therefore it cannot be seen in the
experiments on neutron scattering in the superfluid Helium
[29,30]. However, such possibility cannot be excluded in the
experiments on Raman light scattering. Moreover, in [31], where
such experiments are described, there is the direct indication of
the existence of the gap.

\section*{Acknowledgment}
The authors thank Yu. A. Kuharenko for the useful discussion. V.B.
and S.T. express gratitude to the Netherlands Organization for
Scientific Research (NWO) for support of their investigations on
the problems of statistical physics.

\end{document}